\documentclass[12pt]{iopart}

\eqnobysec

\usepackage[colorlinks=true, pdfstartview=FitV, 
           linkcolor=blue, citecolor=blue, urlcolor=green]{hyperref}
 
\newcommand{\be}{\begin{equation}}
\newcommand{\bear}{\begin{eqnarray}}
\newcommand{\ee}{\end{equation}}
\newcommand{\ear}{\end{eqnarray}}
\renewcommand{\d}{{\rm d}}
\newcommand{\half}{\sfrac12}
\newcommand{\sfrac}[2]{\mbox{$\frac{#1}{#2}$}}
\newcommand{\av}[1]{\left\langle{#1}\right\rangle}
\newcommand{\artanh}{{\rm artanh}}
\newcommand{\bel}[1]{\be\label{#1}
			}
\newcommand{\bearl}[1]{\bear\label{#1}
			}
\newcommand{\req}[1]{(\ref{#1})}
\newcommand{\intx}{\!\int\!}
\newcommand{\overx}[1]{\overline{\phantom{I^o}\!\!\!\!\!#1}}

\begin{document}

\title{Dynamics of an Ising Spin Glass on the Bethe Lattice}

\author{Martin Kiemes${}^{1,2}$ and Heinz Horner${}^{1}$}

\address{${}^1$ Inst. f. Theoretical Physics, Uni Heidelberg,
Philosophenweg 19, D-69120 Heidelberg, Germany}
\address{${}^2$ now at the Inst. f. Theoretical Physics, Uni G\"ottingen,  
   Friedrich-Hund-Platz 1,  D-37077 G\"ottingen, Germany}
\ead{kiemes@theorie.physik.uni-goettingen.de \ horner@tphys.uni-heidelberg.de}
\begin{abstract}
We study the dynamical low temperature behaviour of the Ising spin glass on the Bethe lattice. Starting from Glauber dynamics we propose a cavity like Ansatz that allows for the treatment of the slow (low temperature) part of dynamics. Assuming a continuous phase transitions and ultrametricity with respect to long time scales we approach the problem perturbatively near the critical temperature. The theory is formulated in terms of correlation-response-functions of arbitrary order. They can, however, be broken down completely to products of pair functions depending on two time arguments only. For binary couplings 
$J=\pm I$ a spin glass solution is found which approaches the corresponding solution for the SK-model in the limit of high connectivity. For more general distributions $P(J)$ no stable or marginal solution of this type appears to exist. The nature of the low temperature phase in this more general case is unclear.
\end{abstract}
\submitto{\JPA}
%
%
\section{Introduction}
Most of the work on spin glasses and other systems with frozen-in disorder is based on the evaluation of the free energy or the ground state energy. This comprises replica theory, the cavity method or the TAP-equations \cite{MPV87}. As an alternative stochastic dynamics \cite{SCM04} has been employed for systems with continuous freezing transition \cite{SZ82,Ho84}, e.g. the Sherrington-Kirkpatrick (SK) model \cite{SK75}, and also for systems with discontinuous transition, e.g. the spherical spin glass with $p$-spin interactions \cite{KT87,CHS93,CK93}. The essential difference between the two approaches shows up in the thermodynamic limit. The computation of the free energy does not rely on any kind of dynamics and the question of how equilibrium states can be reached starting from non-equilibrium initial conditions is not adressed. In fact diverging barriers build up in the thermodynamic limit $N\to\infty$. As a consequence the approach via dynamics in the non-ergodic low temperature phase requires regularization by some long time scale, for instance the waiting time $t_w$ after a quench from high temperature, by introducing slowly varying bonds \cite{Ho84} or slow cooling \cite{FH90}. Typically the thermodynamic and long time limit are taken in the order $\lim_{t_w\to\infty} \; \lim_{N\to\infty}$. This means that diverging barriers can not be overcome or the system might be stuck in metastable states \cite{CK93,BBM96}. The importance of the order in which the two limites are performed has been demonstrated in a different context (learning in a perceptron with binary synapses) \cite{Ho92}. For systems with discontinuous transition the dynamic freezing transition is higher than the temperature where replica symmetry breaking sets in, whereas both temperatures are identical for continuous transitions. In this latter case the coefficients of an expansion of the energy or the Edwards-Anderson order parameter near $T_c$ are identical at least up to fourth order \cite{FH90}. At zero temperature, however, the energy obtained within dynamics is expected to be higher than the ground state energy determined within replica theory. Furthermore the energy found within dynamics may depend on details of the cooling schedule \cite{Ho07}.

Spin glasses on diluted graphs, for instance on the Bethe lattice, share finite connectivity with spin glasses in finite dimensions. Nevertheless they are of mean field character as the fully connected SK-model \cite{SK75,MPV87}. Following earlier attempts \cite{Mo87,DG89}, the Ising spin glass on the Bethe lattice has been solved by M\'ezard and Parisi \cite{MP01} using the cavity method on the level of one step replica symmetry breaking. The cavity method benefits from the local tree like structure of the Bethe lattice and deals on the level of 1RSB with distributions of local fields and distributions of those distributions. The resulting functional equations have been solved numerically with a population dynamics algorithm for binary couplings $J=\pm I$.

Investigating the dynamics of systems of Ising spins one has to specify not only the Hamiltonian (energy) but also the kind of dynamics. In the present approach Glauber dynamics, i.e. stochastic single spin flip dynamics with transition probabilities depending on temperature, is used. In the non-ergodic low temperature phase a separation of short and long time scales is assumed. The longest time scale is realized by a waiting time $t_w$ or by some other means \cite{Ho84,FH90}. 
On the short time scale the validity of fluctuation-dissipation theorems are assumed to hold. They are characteristic for equilibrium and in this context they describe equilibrium within a single valley of the energy landscape. On the long time scale in analogy to the cavity method \cite{MP01} a functional equation for the distribution of histories of local fields is derived, investigating the iterative assembly of local subtrees of the Bethe lattice to new subtrees. 

Hierarchically connected equations are derived from this functional for correlation-response-functions of, in principle, arbitrary order. In order to truncate this system of equations, an expansion in powers of the deviation of the temperature from the freezing temperature is employed. Such an expansion is known for the dynamics of the SK-model. 
It turns out that correlation-response-functions of higher order can be broken down completely to expressions involving pair functions depending on two time arguments only. For binary couplings $J=\pm I$ a marginal solution resembling the one for the SK-model is found  \cite{SZ82,Ho84}. 

The situation for more general distributions $P(J)$ is unclear. There exists a critical temperature where the ergodic high temperature solution becomes unstable. The above solution does, however, not apply and no other stable or marginal solution could be found. The expansion contains 
contributions corresponding to the one step replica symmetry breaking solution. 
The original static 1RSB solution \cite{MP01} has been evaluated for $J=\pm I$ only. It would be interesting to see whether this solution persists for general $P(J)$.

The paper is organized as follows. Sec.~2 contains a brief survey of Glauber dynamics and fixes some notations. In Sec.~3 the dynamics of the Ising spin glass on the Bethe lattice is formulated, an effective single site evolution is introduced and equilibrium properties are investigated. Dynamics on long time scales is dealt with in Sec.~4 including a formulation in terms of  distributions of slow components of local fields. In view of an expansion valid in the neighborhood of the freezing temperature, this is reformulated in terms of correlation-response functions. Sec.~5 contains the expansion around the critical temperature in leading order for general $P(J)$ and in next to leading order for $J=\pm I$. The paper concludes with a stability analysis in Sec.~6 and a brief discussion.

\section{Glauber dynamics of a single spin}
The present investigation is based on Glauber dynamics for Ising spins. We start with a brief outline of the formulation \cite{Ho92} describing the dynamics of a single spin. The Hamiltonian of the spin in an external field $h$ is
\be
H=-h\sigma_z.
\ee
The actual state of the spin at time $t$ is described by a 2-component vector
\be
\Big|\rho(t)\Big)=\left(\begin{array}{c}p_{+}(t)\\p_{-}(t)\end{array}\right)=
\left(\begin{array}{c}\half(1+m(t))\\ \half(1-m(t))\end{array}\right).
\ee
Introducing the Pauli matrix
\be
\sigma_z=\left(\begin{array}{cc}1&0\\0&-1\end{array}\right)
\ee
the expectation value of the spin is given by
\be
\av{\sigma_z}=
\begin{array}{cc}(\,1\,&1\,)\\ & \end{array}
\left(\begin{array}{cc}1&0\\0&-1\end{array}\right)
\left(\begin{array}{c}p_{+}(t)\\p_{-}(t)\end{array}\right)=
\Big(\,1\,\Big|\,\sigma_z\,\Big|\rho(t)\Big)=m(t) .
\ee
It is convenient to use a notation resembling quantum mechanics, although Glauber dynamics is purely classical. The time evolution is described by a Liouville operator, a two by two matrix, acting on the state
\be
\partial_t \Big|\rho(t)\Big)
={\cal L}(t)\Big|\rho(t)\Big)
\ee
with Liouvillian
\bel{bead}
{\cal L}(t)
=-\half\gamma
\left(\begin{array}{rr}1-\tanh(\beta h(t))&-1-\tanh(\beta h(t))\\
-1+\tanh(\beta h(t))&1+\tanh(\beta h(t))\end{array}\right) .
\ee
In equilibrium, for constant $h$, ${\cal L}|\,\bar\rho\,)=0$. This determines the equilibrium state
\be
\Big|\,\bar\rho\,\Big)
=\half\left(\begin{array}{c}1+\tanh(\beta h)\\1-\tanh(\beta h)\end{array}\right)
=\frac{\e^{-\beta h \sigma}}{2\cosh(\beta h)} \Big| \;1\;\Big)
\ee
with $|\,1\,)$ being the two dimensional unit vector.
The temporal evolution of the spin is
\be
\dot\sigma_z(t)=\sigma_z {\cal L}(t)-{\cal L}(t)\sigma_z .
\ee
In addition to the spin operator a response operator $\hat \sigma_z(t)$ is introduced. It describes the action of a variation $\delta h(t)$  of the external field at time $t$
\be
\hat \sigma_z=\frac{\partial {\cal L}(t)}{\partial h(t)} .
\ee
Acting on an equilibrium state these operators obey a fluctuation-dissipation theorem (FDT)
\be
\hat\sigma_z\Big|\,\bar\rho\,\Big)= \beta \,\dot\sigma_z\Big|\,\bar\rho\,\Big).
\ee
For time dependent field $h(t)$ the Liouvillians for different times do not commute in general. This can formally be overcome introducing a time ordering operator ${\cal T}$ rearranging products of objects at different time such that 
they are ordered from left to right according to decreasing time. This allows to write for $t>t_0$
\be
\Big|\rho(t)\Big)={\cal T}\Big[\e^{\int_{t_0}^{t}\d t' {\cal L}(t')}\Big]
\Big|\rho(t_0)\Big).
\ee

\section{Ising model on a Bethe lattice}
\subsection{Dynamics}

We consider Ising spins $\sigma_i$ on a Bethe lattice with couplings $J_{ij}$. The Hamiltonian (energy) is
\be
H=-\half\sum_{i,j}J_{ij}\sigma_i\sigma_j-\sum_i h_i\sigma_i .
\ee
The couplings $J_{ij}$ are independent stochastic variables taken from some distribution $P(J)$ such that $\overline{J_{ij}}=0$ and 
$\overline{J_{ij}^2}=I^2$. The underlying vector space for the complete system is spanned by the direct product of the two dimensional vectors associated with each of the spins $\sigma_i$. 

The dynamics is ruled by the Liouvillian
\be
{\cal L}(t)
=\sum_i{\cal L}_i(k_i)
\ee
where ${\cal L}_i(k_i)$ given by \req{bead} acts on the corresponding subspace of site $i$ and depends on the effective field
\bel{fgbm}
k_i=-\frac{\partial H}{\partial \sigma_i}=h_i+\sum_jJ_{ij}\sigma_j .
\ee
Expectation values of time dependent observables with initial condition 
at $t_0$ are written as  
\bearl{fabz}
\fl
\av{A(t) B(t')\Big.\cdots}=\Big(\,1\, \Big |_{\!N\!}A(t) \,
{\cal T}\Big[\e^{\int_{t'}^{t}\d s \sum_i{\cal L}_i(k_i(s))}\Big]\,
B(t')\, {\cal T}\Big[\e^{\int_{\cdots}^{t'}\d s \sum_i{\cal L}_i(k_i(s))}\Big]
\nonumber\\
\cdots {\cal T}\Big[\e^{\int_{t_0}^{\cdots}\d s \sum_i{\cal L}_i(k_i(s))}\Big]
\Big |\,\rho(t_0) \Big)_{\!\!N}
\ear
for  $t>t'>\cdots>t_0$. 

Adopting in the following Ito-calculus the effective fields \req{fgbm} are retarded, i.e. 
\bel{stim}
k_i(t)=h+\sum_j J_{ij}\sigma_j(t^-) .
\ee 
It is convenient to integrate over the effective fields $k_i(t)$ in \req{fabz} and to take \req{stim} into account by introducing appropriate $\delta$-functions. Those are written in their Fourier representation integrating in addition over imaginary auxiliary fields $\hat\kappa_i(t)$. This leads to a path integral representation
\bearl{kjza}
\fl
\av{A(t) B(t')\Big.\cdots}=\prod_i \intx{\cal D}\{\hat k_i, k_i\}\,
\e^{\sum_i\int_{t_0}\!\d s\,\hat k_i(s)\,\{h-k_i(s)\}}
\nonumber\\
\times \! \Big(\,1\, \Big |_{\!N}{\cal T} \bigg[A(t) B(t')\cdots\,
\e^{\sum_{i}\int_{t_0}\!\d s\,\{\hat k_i(s)\sum_jJ_{ij}\sigma_j(s^-)+
{\cal L}_i(k_i(s),s)\}} \bigg] \Big |\,\rho(t_0) \Big)_{\!N}.
\ear

A state $\big |\,\rho\, \big)_{\!N}$ of the system with $N$ sites is in general a superposition of direct products of two dimensional vectors each representing a single spin. The unit vector for $N$ sites $|\,1\,)_{\!N}$ is a direct product of single site unit vectors. $(\,1\,|_{\!N}$ is its adjoint.\\
The equilibrium state, for given $J$, can be expressed as
\be
\Big |\,\bar\rho\, \Big)_{\!N}=Z^{-1}\,\e^{\beta\sum_ih_i\sigma_i+\frac12\beta \sum_{ij}J_{ij}\sigma_i\sigma_j}
\, \Big |\,1\, \Big)_{\!N} .
\ee
It obeys 
\be
{\cal L}_i(k_i) \Big |\;\bar\rho\, \Big)_{\!N}=0 .
\ee

\subsection{Effective single site evolution}

For time dependent expectation values involving the spin at a single site, say site $o$ only, \req{kjza} can be rewritten introducing an effective retarded time evolution. It is obtained by performing the expectation values with respect to all spins except $\sigma_o$. Assuming equilibrium initial conditions expectation values of quantities at a single site, say $o$, can be written as 
\req{kjza}, 
\bearl{dvum}
\fl
\av{\sigma_o(t)\cdots}=\intx{\cal D}\{\hat k, k\}\,
\e^{\int_{t_0}\!\d s\,\hat k(s)\,\{h-k(s)\}}
\nonumber\\
\times \Big(\,1\,\Big |_{o}{\cal T} \Big[\sigma_o(t)\cdots 
{\cal Y}_o(\{\hat k\}\{\sigma_o\};t_0)
\,\e^{\int_{t_0}\!\d s\,{\cal L}_o(k(s),s)}\Big]
\Big|\,\bar\rho_o(k(t_0)) \Big)_{\!o}.
\ear
The action of the remaining part of the system is contained in 
\bearl{rbso}
\fl
{\cal Y}_o(\{\hat k\}\{\sigma\};t_0)=
\prod_i{\Big.}' \intx{\cal D}\{\hat k_i, k_i\}\,
\e^{\sum_i'\int_{t_0}\!\d s\,\hat k_i(s)\,\{h-k_i(s)\}}
\nonumber\\
\fl \quad
\times
\bigg(\,1\, \bigg |_{N-1}{\cal T} \bigg[
\e^{\sum_i'J_{oi}\int_{t_0}\!\d s\,[\hat k(s)\sigma_i(s^-)+
\sigma(s^-)\hat k_i(s)]}\,
\e^{\sum_i'\int_{t_0}\!\d s\,\{\hat k_i(s)\sum_j'J_{ij}\sigma_j(s^-)+
{\cal L}_i(k_i(s),s)\}} \bigg] \bigg |\,\bar\rho\, \bigg)_{\!\!N-1}
\nonumber\\
\fl \quad
=
\prod_i{\Big.}' \intx{\cal D}\{\hat k_i, k_i\}\,
\e^{\sum_i'\int_{t_0}\!\d s\,\hat k_i(s)\,\{h-k_i(s)\}}
\av{\e^{\sum_i'J_{oi}\int_{t_0}\!\d s\,[\hat k(s)\sigma_i(s^-)+
\sigma(s^-)\hat k_i(s)]}
\Bigg.}_{\!\!N-1}
\ear
and the initial condition in \req{dvum} is
\be
\Big|\bar\rho_o(k)\Big)_{\!o}=\frac1{2\cosh(\beta k)}\,\e^{\beta k \sigma_o}
\Big|\,1\,\Big)_{\!o} .
\ee 
In the above expression the primed products and sums run over $i\ne o$. 
The expectation values in \req{rbso} refer to the system without the spin at site $o$.

It should be pointed out that the above expression holds for given values of the couplings $J_{ij}$ and the average over $J_{ij}$ has still to be evaluated.

\subsection{Equilibrium}

The above path integral comprises an integration at the initial time $t_0$. This is investigated in the following. The single time (static) expectation value of $\sigma_o(t_o)$ is
\be
\av{\sigma_o(t_0)\Big.}
=Z_o^{-1}\intx \frac{\d \hat k \d k}{2\pi}\,\e^{\hat k\,\{h-k\}}
(\,1\,| \,\sigma\,Y_o(\hat k)\,\e^{\beta k \sigma}|\,1\,)
\ee
and the effective action \req{rbso} reduces to
\be
Y_o(\hat k)=\av{\e^{\hat k\sum_i' J_{oi}\sigma_i}}_{\!\!N-1} .
\ee
In analogy to the cavity method we define $c_o$ and $k_o$ writing
\be
\e^{\beta(c_o+ k_o\sigma)}=
\intx \frac{\d \hat k \d k}{2\pi}\,\e^{\hat k\,\{h-k\}}
\,Y_o(\hat k)\,\e^{\beta k \sigma} .
\ee
This yields for the partition function and the local magnetization
\bel{dvkq}
Z_o=2\,\e^{\beta c_o}\cosh(\beta k_o)
\qquad \mbox{and}\qquad
\av{\sigma_o\big.}=\tanh(\beta k_o)
\ee
The two quantities defined above result in
\be
c_o=\frac1\beta\Big\{\ln\Big(
\av{\e^{\beta\sum_i' J_{oi}\sigma_i}\Big.}_{\!\!N-1}\Big)+
\ln\Big(\av{\e^{-\beta\sum_i' J_{oi}\sigma_i}\Big.}_{\!\!N-1}\Big)\Big\}
\ee
\be
k_o=h+\frac1{2\beta}\Big\{\ln\Big(
\av{\e^{\beta\sum_i' J_{oi}\sigma_i}\Big.}_{\!\!N-1}\Big)-
\ln\Big(\av{\e^{-\beta\sum_i' J_{oi}\sigma_i}\Big.}_{\!\!N-1}\Big)\Big\} .
\ee

For the SK-model and for the Bethe lattice the expectation values for sites $i$ can be assumed to factorize. For Ising spins and using \req{dvkq} we find the identity
\be
\av{\e^{\pm\beta J_{oi}\sigma_i}\Big.}_{\!\!N-1}=
\cosh(\beta J_{oi})\pm \tanh(\beta k_i)\sinh(\beta J_{oi})
\ee
which allows to express $c_o$ and $k_o$ in terms of the fields of the adjacent sites:
\bel{badw}
c_o=\sum_i \Big.'c(J_{oi},k_i) , \qquad\quad 
k_o=h+\sum_i\Big.' u(J_{oi},k_i) ,
\ee
\be
c(J,k)=\sfrac1{2\beta}\,\ln\big(\cosh^2(\beta J)
-\tanh^2(\beta k) \sinh^2(\beta J)\big),
\ee
\bel{srfe}
u(J,k)=\sfrac1\beta \, \artanh\big(\tanh(\beta k)\tanh(\beta J)\big).
\ee
In particular for $J\to0$, i.e. for the SK-model,
\be
u(J,k)=J\,\tanh(\beta k) \qquad\qquad
c(J,k)=\frac{\beta\,J^2}{2\,\cosh^2(\beta k)} .
\ee
In view of this limit correlation-functions are later defined with rescaled quantities
\bel{dbtl}
U(J,\kappa)=\sfrac1I u(J,\kappa) .
\ee
where $I$ is the typical size  of the couplings. This ensures e.g. that the Edwards-Anderson order parameter stays finite in the SK-limit.

\section{Slow Dynamics}
\subsection{Distribution of slow fields}
Dynamics on short time scales takes place within a single valley of the energy landscape, dynamics on long time scales is supposed to be due to transitions among different valleys. Assume there exists a time scale $t^* \gg 1$ separating fast and slow motions. This time scale might be realized by some waiting time, slowly changing couplings or other means \cite{Ho84,FH90}.
The local field $k(t)$ and the conjugate field $\hat k(t)$ are split into fast and a slow contributions
\be
k(t)\to k(t)+\kappa(t/t^*), \qquad\qquad
\hat k(t)\to\hat k(t)+\frac{1}{t^*}\hat\kappa(t/t^*).
\ee
The fast parts are due to fluctuations around the quasi equilibrium state within 
a single valley. The slow part $\kappa(\tau)$ acts like an external field and the quasi equilibrium state follows this field adiabatically.
Evaluating time dependent correlation functions of the spin $\sigma_0$ on this long time scale, this spin can be viewed as being in equilibrium in the slow field resulting in
\bel{dolt}
\av{\sigma_o(t^* \tau)\,\sigma_o(t^* \tau')\cdots}
=\tanh\big(\beta\kappa_o(\tau)\big)\tanh\big(\beta\kappa_o(\tau')\big)
\cdots .
\ee
The effective field is given by \req{badw} with $k$ replaced by 
$\kappa(\tau)$, i.e.
\bel{usfr}
\kappa_o(\tau)=h+\sum_i\Big.' u(J_{oi},\kappa_i(\tau^-)).
\ee
In the low temperature regime of a spin glass this field is, however, distributed and \req{dolt} has to be averaged over some distribution 
${\cal P}_o(\{\kappa_0\})$. With \req{usfr} this distribution can be calculated from the distribution of slow fields at the neighboring sites $i$. Continuing this mapping to outer shells of the lattice an iteration scheme is set up similar to the course of action in the cavity approach \cite{MP01}. This program involves the following steps:

Performing the average over the couplings $J_{ij}$ 
and defining
\bel{ljtw}
{\cal Q}_o(\{\hat\kappa,\kappa\})=\overline{\av{\e^{\!\int\!\d\tau\sum_i'
\{\hat\kappa(\tau) u(J_{oi},\kappa_i(\tau^-))+
\hat\kappa_i(\tau)u(J_{oi},\kappa(\tau^-))\}}\bigg.}}^J_{\!\!N-1}
\ee
this distribution is obtained by integration over $\hat\kappa$ 
\bel{duhw}
{\cal P}_o(\{\kappa\})=\intx{\cal D}\{\hat\kappa\}\,
\e^{\int\!\d\tau \,\hat\kappa(\tau)\{h-\kappa(\tau)\}}
{\cal Q}_o(\{\hat\kappa,\kappa\}) .
\ee
The average on the right hand side of \req{ljtw} involves a corresponding joint distribution ${\cal Q}_{N-1}$ of the fields $\hat\kappa_i$ and $\kappa_i$ at the $K+1$ neighboring sites. The second term in the exponent is the contribution of site $o$ to the local field at site $i$, corresponding to the Onsager reaction field.

Assuming factorization of the expectation values on the right hand side of
\req{ljtw} leads to
\bearl{ljuy}
\fl
{\cal Q}_o(\{\hat\kappa,\kappa\})=
\prod_i\Big.'\!\intx {\cal D}\{\hat\kappa_i,\kappa_i\}\,
\e^{\int\!\d\tau\,\hat\kappa_i(\tau)\{h-\kappa_i(\tau)\}}
{\cal Q}_i(\{\hat\kappa_i,\kappa_i\})
\nonumber\\
\times
\overline{\e^{\!\int\!\d\tau\{\hat\kappa(\tau) 
u(J_{oi},\kappa_i(\tau^-))
+\hat\kappa_i(\tau)u(J_{oi},\kappa(\tau^-)) \}}}^{J_{oi}} .
\ear
The functional ${\cal Q}_o(\{\hat\kappa,\kappa\})$ contains the action of all 
$K+1$ spins surrounding site $o$. ${\cal Q}_i(\{\hat\kappa_i,\kappa_i\})$, on the other hand, contains only the action of the $K$ spins on the outgoing branches originating at site $i$. The action of the spin $\sigma_o$ is taken into account by the bond averaged exponential in \req{ljuy}. The functional 
${\cal Q}_i$ is given by a similar average over the distributions on the next shell of sites on the tree. Iterating this process, a fixed point distribution is assumed to exist. It is given by
\bearl{stgq}
\fl
{\cal Q}(\{\hat\kappa,\kappa\})=
\bigg[\int\!{\cal D}\{\hat \lambda,\lambda\}\,
\e^{\int\!\d\tau\,\hat\lambda(\tau)\{h-\lambda(\tau)\}}\,
{\cal Q}(\{\hat\lambda,\lambda\})
\nonumber\\
\times
\overline{\e^{\!\int\!\d\tau\{\hat\kappa(\tau) 
u(J,\lambda(\tau^-))
+\hat\lambda(\tau) u(J,\kappa(\tau^-))\}}}^{J}\,\bigg]^K .
\ear
The functional ${\cal Q}_o(\{\hat\kappa,\kappa\})$ for the central spin at site $o$ is given by the same expression with $K$ replaced by $K+1$.

The above line of arguments follows pretty much the steps in the cavity method calculation of M\'ezard and Parisi \cite{MP01}. The functional fixed point equation \req{stgq} is, however, more general. The only assumptions made are the separation of time scales and the factorization of expectation values on different subtrees. In particular it contains in principle contributions corresponding to full replica symmetry breaking solutions. 
Within this formulation the replica symmetric solution is obtained by leaving out the reaction term in \req{stgq} as shown later.
\subsection{Correlation-response-functions}
There is little hope to find solutions without relying on approximations or expansions. In the following we investigate an expansion in powers of the difference $\epsilon=\frac{T_c-T}{T_c}$ between transition temperature $T_c$ and the actual temperature $T$. Such an expansion is expected to apply in the neighborhood of a continuous freezing transition where the Edwards-Anderson order parameter $q_{EA}\sim\epsilon$ near $T_c$. This is the case for the SK-model. 

Rather than working with the full functional ${\cal Q}(\{\hat\kappa,\kappa\})$, it is sufficient to investigate its moments, the correlation-response-functions on the long time scale. With the rescaled quantities $U(J,\kappa)$, \req{dbtl},
they are defined as
\bearl{brxo}
\fl
C_{nm}(\tau_1,\cdots,\tau_n;\tau_1',\cdots,\tau_m';J)=
\intx{\cal D}\{\hat\kappa,\kappa\}\,
\e^{\int\!\d\sigma\,\hat\kappa(\sigma)\{h-\kappa(\sigma)\}}\,
{\cal Q}(\{\hat\kappa,\kappa\})
\nonumber\\
\times \prod_{\nu=1}^n U(J,\kappa(\tau_\nu))
\prod_{\mu=1}^m \hat\kappa(\tau_\mu') .
\ear
Inserting \req{stgq} and expanding the exponential, a hierarchy of equations for the correlation-response-functions is obtained which is derived along the following lines: Evaluating a function $C_{nm}(\tau_1,\cdots,\tau_n;\tau_1',\cdots,\tau_m';J)$ there is the product over $U(J,\kappa(\tau_\nu))$ involving $\kappa$ at the times $\tau_\nu$. The factors 
$\hat\kappa(\tau_\mu')$ can be replaced under the path integral by functional derivatives $\delta/\delta \kappa(\tau_\mu)$ acting on ${\cal Q}(\{\hat\kappa,\kappa\})$, at least as long as all time arguments are different (otherwise the derivatives could act on the products of $U(J,\kappa(\tau_\nu))$ as well). The derivatives acting on ${\cal Q}$ create additional terms involving $\kappa(\tau_\mu')$. Expanding the second term in the exponential of \req{stgq}, internal integrations over times $\sigma_\rho$ are generated. They contribute additional factors now depending on $\kappa(\sigma_\rho)$. In order $l$ of this expansion there are alltogether functions of $n+m+l$ discrete times. 
For $\tau\ne \tau_\nu,\tau_\mu',\sigma_\rho$ the path integral over $\kappa(\tau)$ can be performed resulting in $\hat\kappa(\tau)=0$. This leaves integrations over $\hat\kappa$ and $\kappa$ at those discrete times only. With \req{stgq} they reduce to functions of correlation-response-functions. This general strategy is applied in the following to correlation-response-functions of lowest order.

\section{Expansion around $T_c$}
\subsection{Leading order for general $P(J)$}\label{FNU}

We investigate an expansion in powers of $\epsilon=\frac{T_c-T}{T_c}$. For the SK-model $C_{nm}\sim \epsilon^{(n+3m)/2}$ is found \cite{SZ82,Ho84} and the same scaling with $\epsilon$ is proposed for the present investigation as well. Later it is shown that this scaling is fulfilled in a consistent manner.
For simplicity the following discussion will be restricted to $h=0$.

We may start with an expansion of \req{stgq}
\bearl{ehmp}
\fl
{\cal Q}(\{\hat\kappa,\kappa\})\approx
\bigg[1+\half I^2\!\intx\d\sigma\d\sigma'\,
\hat\kappa(\sigma)\hat\kappa(\sigma')\,
\overline{C_{2,0}(\sigma,\sigma';J')}^{J'}
\nonumber\\
+I^2\!\intx\d\sigma\d\sigma'\,\hat\kappa(\sigma)\,
\overline{U(J',\kappa(\sigma'))\,C_{1,1}(\sigma,\sigma';J')}^{J'}
\nonumber\\
+\sfrac{1}{4!}I^4 \!\intx\d\sigma_1\d\sigma_2\d\sigma_3\d\sigma_4\,
\hat\kappa(\sigma_1)\hat\kappa(\sigma_2)\hat\kappa(\sigma_3)\hat\kappa(\sigma_4)\,\overline{C_{4,0}(\sigma_1,\sigma_2,\sigma_3,\sigma_4;J')}^{J'}
\nonumber\\
+\sfrac{1}{3!}I^4\! \intx\d\sigma_1\d\sigma_2\d\sigma_3\d\sigma_4\,
\hat\kappa(\sigma_1)\hat\kappa(\sigma_2)\hat\kappa(\sigma_3)\,
\overline{U(J',\kappa(\sigma_4))\,C_{3,1}(\sigma_1,\sigma_2,\sigma_3;\sigma_4;J')}^{J'}
\nonumber\\
+\cdots \bigg]^K .
\ear
Evaluating $C_{2,0}(\tau,\tau';J)$ and keeping terms $\sim\epsilon^2$
\bearl{gavb}
\fl
C_{2,0}(\tau,\tau';J)=\!\int\!{\cal D}\{\hat\kappa,\kappa\}\,
\e^{\int\!\d\sigma\,\hat\kappa(\sigma)\{h-\kappa(\sigma)\}}
U(J,\kappa(\tau))U(J,\kappa(\tau'))
\nonumber\\
\bigg\{1+\half I^2 K \! \intx\d\sigma\d\sigma'\,
\hat\kappa(\sigma)\hat\kappa(\sigma')\,\overline{C_{2,0}(\sigma,\sigma';J')}^{J'}
\nonumber\\
+\sfrac{1}{4!} I^4 K\!\intx\d\sigma_1\d\sigma_2\d\sigma_3\d\sigma_4\,
\hat\kappa(\sigma_1)\hat\kappa(\sigma_2)\hat\kappa(\sigma_3)\hat\kappa(\sigma_4)\\
\times
\bigg[\overline{C_{4,0}(\sigma_1,\sigma_2,\sigma_3,\sigma_4;J')}^{J'}
+3(K-1) \overline{C_{2,0}(\sigma_1,\sigma_2;J')}^{J'}\,
\overline{C_{2,0}(\sigma_3,\sigma_4;J'')}^{J''}\bigg]\bigg\} .
\nonumber
\ear
Performing the path integration over $\kappa(\sigma)$ for $\sigma\ne\tau,\tau'$ the integrals over $\sigma_{\!..}$ reduce to sums over 
$\sigma_{\!..}=\tau,\tau'$ and the path integrals become ordinary integrations over variables $\hat\kappa=\hat\kappa(\tau)$, $\kappa=\kappa(\tau)$, 
$\hat\kappa'=\hat\kappa(\tau')$ and $\kappa'=\kappa(\tau')$.

It is convenient to introduce vertices
\bel{kehn}
V_n(J)=\frac{\partial^{n-1}}{\partial \kappa^{n-1}} U(J,\kappa)\big|_{\kappa=0}
\ee
with $V_n=0$ for odd $n$. In the following we use the notation 
$\overx{A(J)B(J)}^J\to \overx{AB}$. 
Then \req{gavb} yields
\bear
\fl
C_{2,0}(\tau,\tau';J)=\bigg\{I^2 K V_2^2(J)+\half I^4 K(K-1) V_2(J)V_4(J)
\Big[\,\overx{C_{2,0}}(\tau,\tau)
+\overx{C_{2,0}}(\tau',\tau')\,\Big]\bigg\} \,
\nonumber\\
\times 
\overx{C_{2,0}}(\tau,\tau')
+\sfrac{1}{3!}I^4 K\,V_2(J)V_4(J)
\Big[\,\overx{C_{4,0}}(\tau,\tau,\tau,\tau')
+\overx{C_{4,0}}(\tau,\tau',\tau',\tau')\,\Big] .
\ear
Evaluation of $C_{4,0}$, again with $h=0$, in order $\epsilon^2$ results in
\bear
\fl
C_{4,0}(\tau_1,\tau_2,\tau_3,\tau_4;J)=I^4 K V_2^4(J)\,
\overx{C_{4,0}}(\tau_1,\tau_2,\tau_3,\tau_4)
+I^4 K (K-1) V_2^4(J) 
\\
\times\!\Big[\,\overx{C_{2,0}}(\tau_1,\tau_2)\,
\overx{C_{2,0}}(\tau_3,\tau_4)
\!+\!\overx{C_{2,0}}(\tau_1,\tau_3)\,
\overx{C_{2,0}}(\tau_2,\tau_4)
\!+\!\overx{C_{2,0}}(\tau_1,\tau_4)\,
\overx{C_{2,0}}(\tau_2,\tau_3)\,\Big]
\nonumber
\ear
and performing the average over $J$ we get
\bear
\fl
\overx{C_{4,0}}(\tau_1,\tau_2,\tau_3,\tau_4)=
I^4 K(K-1)\frac{\overx{V_2^4}}{1-I^2 K \overx{V_2^4}}
\Big[\,\overx{C_{2,0}}(\tau_1,\tau_2)\;
\overx{C_{2,0}}(\tau_3,\tau_4)
\nonumber\\
+\overx{C_{2,0}}(\tau_1,\tau_3)\;
\overx{C_{2,0}}(\tau_2,\tau_4)
+\overx{C_{2,0}}(\tau_1,\tau_4)\,
\overx{C_{2,0}}(\tau_2,\tau_3)\,\Big] .
\ear
With this $C_{2,0}$ becomes
\bearl{fznw}
\fl
C_{2,0}(\tau,\tau';J)=\bigg\{I^2 K V_2^2(J)+\half I^4 K(K-1)
\frac{V_2(J)V_4(J)}{1-I^4 K \overx{V_2^4}}
\Big[\,\overx{C_{2,0}}(\tau,\tau)
+\overx{C_{2,0}}(\tau',\tau')\,\Big]\bigg\}
\nonumber\\
\times\overx{C_{2,0}}(\tau,\tau') .
\ear
It should be noted that the correlation function $\overx{C_{4,0}}$ has been broken down completely to products of pair functions $\overx{C_{2,0}}$. This can be done for correlation-response-functions $\overx{C_{n,m}}$ with general $n$ and $m$ in higher orders as well. In this case the products contain also response functions $\overx{C_{1,1}}$. 

Averaging over $J$ a non trivial solution with $\overx{C_{2,0}}(\tau,\tau')\ne 0$ requires that the expression in the curly bracket on the right hand side equals $1$. Including terms of higher orders in $\epsilon$ shows that a corresponding expression holds only for the time derivative. Assuming time translational invariance we can write
\bel{lkzy}
\partial_\tau\overx{C_{2,0}}(\tau)=Y(\tau)\,\partial_\tau\overx{C_{2,0}}(\tau)
\ee
with
\bel{dvft}
Y(\tau)=I^2 K \overx{V_2^2}+I^4K(K-1)
\frac{\overx{V_2 V_4}}{1-I^4 K \overx{V_2^4}}
\overx{C_{2,0}}(0) .
\ee
In first order $Y(\tau)$ is constant and \req{lkzy} would hold for 
$\overx{C_{2,0}}(\tau)$ as well. Including higher orders, however,  it depends explicitly on $\tau$ and \req{lkzy} holds for the derivative only.
A non trivial solution requires $Y(\tau)=1$ for all $\tau$ and this eventually  determines the full time dependence of $\overx{C_{2,0}}(\tau)$.

The evaluation of $C_{1,1}$ follows similar lines. Since 
$C_{1,1}\sim \epsilon^2$ contributions $\sim\epsilon^3$ are kept, and the response function becomes
\bearl{krnz}
\fl
C_{1,1}(\tau,\tau';J)= I^2 K V_2(J)\,
\overx{V_2 C_{1,1}}(\tau,\tau')
+\half I^4 K(K-1) V_4(J)\,\overx{C_{2,0}}(\tau,\tau)\,
\overx{V_2 C_{1,1}}(\tau,\tau')
\nonumber\\
+\half I^4 K(K-1) V_2(J)\,\overx{V_4 C_{1,1}}(\tau;\tau')\,
\overx{C_{2,0}}(\tau',\tau')
\nonumber\\
+\sfrac{1}{3!}I^4 KV_4(J)\,
\overx{V_2 C_{3,1}}(\tau,\tau,\tau;\tau')
+\half I^4 K V_2(J)\,\overx{V_4 C_{3,1}}(\tau,\tau',\tau';\tau') .
\ear
Likewise the four point function on the right hand side obeys
\bearl{vzew}
\fl
C_{3,1}(\tau_1,\tau_2,\tau_3;\tau_4;J)=I^4 K V_2^3(J)\,
\overx{V_2 C_{3,1}}(\tau_1,\tau_2,\tau_3;\tau_4)
\nonumber\\
+I^4 K (K-1) V_2^3(J) \Big[\,\overx{V_2 C_{1,1}}(\tau_1;\tau_4)\;
\overx{C_{2,0}}(\tau_2,\tau_3)
\nonumber\\
+\overx{V_2 C_{1,1}}(\tau_2;\tau_4)\;
\overx{C_{2,0}}(\tau_1,\tau_3)
+\overx{V_2 C_{1,1}}(\tau_3;\tau_4)\;
\overx{C_{2,0}}(\tau_1,\tau_2)\,\Big] .
\ear
The average over $J$ requires some attention because 
on the right hand side of \req{krnz} and \req{vzew} the averages contain different vertices $V_n(J)$. Nevertheless they can be evaluated successively. Multiplying \req{vzew} with $V_2(J)$ and performing the $J$-average,
$\overx{V_2 C_{3,1}}$ can be expressed in terms of $\overx{V_2 C_{1,1}}$ and 
$\overx{V_2 C_{1,1}}$. Multiplying \req{vzew} with $V_4(J)$, averaging and using the result for $\overx{V_2 C_{3,1}}$, $\overx{V_4 C_{3,1}}$ is again written in terms of pair functions. This is inserted into \req{krnz} resulting in
\bearl{strn}
\fl
C_{1,1}(\tau;\tau';J)\!=\!\bigg[I^2 K V_2(J)+\half I^4 K(K\!-\!1)\, 
\frac{V_4(J)\overx{C_{2,0}}(\tau,\tau)
\!+\!I^4 K V_2(J) \overx{V_2^3V_4}\,\overx{C_{2,0}}(\tau',\tau')}{1-I^4 K \overx{V_2^4}}\,\bigg]
\nonumber\\
\times \overx{V_2 C_{1,1}}(\tau;\tau')
+ \half I^4  K(K-1)V_2(J) \overx{C_{2,0}}(\tau',\tau')\,\overx{V_4C_{1,1}}(\tau;\tau') .
\ear
This yields in lowest order
\be
\overx{V_4C_{1,1}}(\tau;\tau')=I^2 K \,\overx{V_2V_4}\,
\overx{V_2C_{1,1}}(\tau;\tau') .
\ee
Inserted into \req{strn} a closed equation for $\overx{V_2C_{1,1}}$ is obtained
\bear
\fl
\overx{V_2C_{1,1}}(\tau;\tau')= \bigg\{I^2 K \overx{V_2^2}+\half I^4 K(K-1)\, 
\frac{\overx{V_2V_4}}{1-I^4 K \overx{V_2^4}}\,
\overx{C_{2,0}}(\tau,\tau)
\\
+ \half I^4 K (K-1)\bigg[I^2 K \overx{V_2^2}\,\overx{V_2V_4}+
\frac{I^4 K \overx{V_2^2}\, \overx{V_2^3V_4}}{1-I^4 K \overx{V_2^4}}\bigg] 
\overx{C_{2,0}}(\tau',\tau')\bigg\}\overx{V_2C_{1,1}}(\tau;\tau')
\nonumber
\ear
which is again of the form
\bel{fyce}
\overx{V_2C_{1,1}}(\tau)=Y'(\tau)\,\overx{V_2C_{1,1}}(\tau)
\ee
assuming time translational invariance. 

Pointing out the difference between $Y(\tau)$, \req{dvft}, and $Y'(\tau)$ we write
\bel{gilr}
\fl
Y'(\tau)=Y(\tau)+\half I^4 K(K\!-\!1) 
\bigg\{\frac{I^4K \big(\overx{V_2^2}\,\overx{V_2^3V_4}
\!-\!\overx{V_2^4}\,\overx{V_2V_4}\big)}
{1-I^4 K\overx{V_2^4}}
-\Big[1\!-\!I^2 K \overx{V_2^2}\Big]\overx{V_2V_4}\bigg\}\overx{C_{2,0}}(0).\;
\ee
This allows for the following non trivial solutions:\\
(a) \ \ $Y(\tau)=1$, \ $Y'(\tau)\ne 1$ : \ $\overx{C_{2,0}}(\tau)\ne0$, \ 
$\overx{C_{1,1}}(\tau)=0 \bigg.$.\\
(b) \ \ $Y(\tau)\ne 1$, \ $Y'(\tau)= 1$ : \ $\overx{C_{2,0}}(\tau)=0$, \ 
$\overx{C_{1,1}}(\tau)\ne0\bigg.$.\\
(c) \ \ $Y(\tau) = Y'(\tau)= 1$ : \qquad $\!\overx{C_{2,0}}(\tau)\ne 0$, \ 
$\overx{C_{1,1}}(\tau)\ne0\bigg.$. 

A solution of type (a) has been found for instance for the spherical version of the SK-model\cite{ZKH00}. It does not describe glassy behavior since the response function vanishes on the long time scale, although the correlation function shows non trivial properties. A solution of type (b) does actually not exist as shown later.

Only the last solution is supposed to be characteristic for a spin glass phase. Since $I^2K \overline{V_2^2}=1+{\cal O}(\epsilon)$ the last term in \req{gilr} does not contribute in order $\epsilon$ and only the first term in the bracket has to be taken into account. This means, however, that solution (c) requires
\bel{xfkm}
\overx{V_2^2}\,\overx{V_2^3V_4}=\overx{V_2^4}\,\overx{V_2V_4} .
\ee
With \req{srfe}, \req{dbtl} and \req{kehn}
\be
\fl
V_2(J)=I^{-1}\tanh(\beta J) \qquad
V_4(J)=-2 \beta^2 I^{-1} \tanh(\beta J) \, \big(1-\tanh^2(\beta J)\big)
\ee
and
\bearl{ntas}
\fl
V_2^2(J)\,V_2^3(J')V_4(J')-V_2^4(J)\,V_2(J')V_4(J')
\nonumber\\
=2\beta^2 I^{-6} \tanh^2(\beta J) \tanh^2(\beta J') 
\big(\tanh^2(\beta J)-\tanh^2(\beta J')\big)^2 .
\ear
This shows that \req{xfkm} can be fulfilled only with $J=\pm I$ and that this kind of solution does not exist for more general $P(J)$. This result is quite remarkable, indicating the special role of the binary distribution of couplings.

\subsection{Next to leading order for $J=\pm I$}

The correlation-response-functions \req{brxo} have the symmetry
\be 
C_{nm}(\cdots;J)=(-1)^n C_{nm}(\cdots;-J).
\ee
For binary couplings $J=\pm I$ the bond average therefore becomes almost trivial. For simplicity we set $h=0$ in the following  and assume time translational invariance on the long time scale. The strategy to evaluate contributions of higher orders in $\epsilon$ is analogous to what has been described in section \ref{FNU}. Evaluating 
$C_{2,0}$ in order $\epsilon^3$ requires to compute $C_{4,0}$ and $C_{6,0}$ in the same order and to break them down successively to products of pair functions 
$C_{2,0}$. In addition in this order an internal integration shows up taking into account the third line on the right hand side of \req{ehmp} leading to a contribution of the form
\be
C_{2,0}(\tau,\tau')=\cdots + V_2 \intx\d\sigma \,
C_{3,1}(\tau,\tau',\sigma;\sigma^+) .
\ee
The four point function $C_{3,1}$ can again be broken down to products of pair functions according to \req{vzew}.

Using the notation
\be
\fl
\bar q(\tau-\tau')=C_{2,0}(\tau,\tau';I), \qquad
\bar r(\tau-\tau')=C_{1,1}(\tau;\tau';I), \qquad
V_n=V_n(I)
\ee
and collecting all contributions up to order $\epsilon^3$ one obtains
\bearl{mfqa}
\fl
\bar q(\tau)=I^2 K\,V_2^2\, \bar q(\tau)
+I^4 K(K-1)\,\frac{V_2V_4 }{1-I^4K\,V_2^4 }\,
\bar q(0)\,\bar {q }(\tau)
\nonumber\\
\fl \quad
+ I^6 K(K-1)\,\frac{1 }{1-I^4K\,V_2^6 }
\bigg\{K-2+3I^4 K(K-1)\frac{V_2^4}{1-I^4 K V_2^4}\bigg\}
\nonumber\\
\fl \quad
\times
\bigg\{\bigg[\sfrac14 \big(V_2V_6\!+\!V_4^2 \big)
+ 3 I^4 K \frac{V_2^4V_4^2 }{1\!-\!I^4K\,V_2^4}\bigg]\bar {q }^2(0)
+\sfrac1{3!}\bigg[V_4^2 + 2 I^4 K \frac{V_2^4V_4^2 }{1\!-\!I^4KV_2^4}\bigg]
\bar {q }^2(\tau) \bigg\}\,
\bar {q }(\tau)
\nonumber\\
\fl \quad
+I^4 K(K-1)\,
\frac{{V_2^3 }}{1-I^4K{V_2^4 }}\,
\!\int\!\d\sigma
\bigg\{{ \bar r }(\tau+\sigma) 
\bar {q }(\sigma)
+\bar {q }(\tau-\sigma) 
{\bar r }(\sigma)\bigg\} .
\ear
The corresponding calculation for the response function yields
\bearl{rkda}
\fl
\bar r(\tau)=I^2 K\,V_2^2\, \bar r(\tau)
+I^4 K(K-1)\,\frac{V_2V_4 }{1-I^4KV_2^4 }\,
\bar q(0)\,\bar r (\tau)
\nonumber\\
\fl \quad
+ I^6 K(K-1)\,\frac{1 }{1-I^6KV_2^6 }
\bigg\{K-2+3I^4K(K-1)\frac{V_2^4}{1-I^4K V_2^4}\bigg\}
\nonumber\\
\fl \quad
\times
\bigg\{\bigg[\sfrac14 \big(V_2V_6\!+\!V_4^2 \big)
+ 3 I^4 K \frac{V_2^4V_4^2 }{1\!-\!I^4KV_2^4}\bigg]\bar {q }^2(0)
+\sfrac12\bigg[V_4^2 + 2 I^4 K \frac{V_2^4V_4^2 }{1\!-\!I^4KV_2^4}\bigg]
 \bar {q }^2(\tau) \bigg\}
\,\bar r(\tau)
\nonumber\\
\fl \quad
+I^4K(K-1)\,
\frac{{V_2^3 }}{1-I^4K\,{V_2^4 }}\,
\!\int\!\d\sigma\,\bar r (\tau-\sigma) \,\bar r (\sigma) .
\ear

\subsection{Ultrametric time parametrization}

The low temperature phase of the SK-model is characterized by a hierarchy of long time scales ranging from $t^*$ to some longest time scale $t_w$. Eventually the limit $t_w\!\to \!\infty$ is taken. In order to keep track of the long time scales it is convenient to introduce the parameterization \cite{Ho84}
\be
t=t_w^{1-x(\tau)}, \qquad\qquad \quad
x(t)=1-\frac{\ln(t)}{\ln(t_w)} ,
\ee
and write
\be
\bar q(\tau)=\bar Q\big(x(t^*\tau)\big), \qquad
\bar r(\tau)=-t^*\,\dot x(t^*\tau)\,\bar R\big(x(t^*\tau)\big) .
\ee
With the above definition
\be
x(t+t')=1-\frac{\ln(t_w^{1-x(t)}+t_w^{1-x(t')})}{\ln(t_w)}
=x(t)-\frac{\ln(1+t_w^{x(t)-x(t')})}{\ln(t_w)}
\ee
and for $t>t'$ and with it $x(t)<x(t')$
\be
x(t+t')=x(t)-\frac{\ln(1+t_w^{x(t)-x(t')})}{\ln(t_w)}\approx
x(t)-\frac{t_w^{x(t)-x(t')}}{\ln(t_w)}\approx x(t) .
\ee
The corresponding result for $t<t$ is $ x(t+t')\approx x(t')$. This yields the ultrametric relation
\be
x(t+t')=x\big(\max(t,t')\big) .
\ee 
With it the integrals in \req{mfqa} and \req{rkda} become
\bear
\fl
\int_0^\tau\!\!\d\sigma\,\bar r(\tau-\sigma)\bar q(\sigma)+\!
\int_0^\infty\!\!\!\d\sigma\,\Big\{\bar r(\tau+\sigma)\bar q(\sigma)+
\bar q(\tau+\sigma)\bar r(\sigma)\Big\}
\nonumber\\
\to 
2 \bar Q(x) \!\int_x^1\!\!\d x'\,\bar R(x')
+2\!\int_0^x\!\!\d x'\,\bar Q(x')\bar R(x')
\ear
and
\be
\int_0^\tau\!\!\d\sigma\,\bar r(\tau-\sigma)\bar r(\sigma)
\to 
-2 t^*\dot x(t*\tau)\, \bar R(x) \!\int_x^1\!\!\d x'\,\bar R(x')
\ee
This allows to write \req{rkda} in the form
\be
\bar R(x)= \bar Y(x)\,\bar R(x)
\ee
with
\bel{hqfc}
\bar Y(x)=Y_0+Y_1\bar Q(1) +\half Y_2 \bar Q^2(1)+\half Y_{2x} \bar Q^2(x)
+Y_r\!\int_x^1\!\!\d x'\,\bar R(x')
\ee
and the coefficients
\bel{dbga}
\fl
Y_0=I^2K\,V_2^2 ,
\ee
\be
\fl
Y_1=I^4K(K-1)\,\frac{V_2V_4 }{1-I^4KV_2^4 } ,
\ee
\bearl{hesd}
\fl
Y_2=I^6K(K-1)\,\frac{1 }{1-I^4K\,V_2^6 }
\bigg\{K-2+3I^4K(K-1)\frac{V_2^4}{1-I^4K V_2^4}\bigg\}
\nonumber\\
\times\bigg\{\sfrac12 \big(V_2V_6+V_4^2 \big)
+ 6 I^4 K \frac{V_2^4V_4^2 }{1\!-\!I^4KV_2^4}\bigg\} ,
\ear
\bear
\fl
Y_{2x}=I^6K(K-1)\,\frac{1 }{1-I^6KV_2^6 }
\bigg\{K-2+3I^4K(K-1)\frac{V_2^4}{1-I^4K V_2^4}\bigg\}
\nonumber\\
\times
\bigg\{V_4^2+ 2 I^4 K \frac{V_2^4V_4^2 }{1-I^4KV_2^4}\bigg\} ,
\ear
and
\bel{kytf}
\fl
Y_r=2 I^4 K(K-1)\,\frac{{V_2^3 }}{1-I^4K\,{V_2^4 }} .
\ee
The correlation function \req{mfqa} becomes
\bear
\fl
\bar Q(x)= \bigg\{Y_0+Y_1\bar Q(1) +\half Y_2 \bar Q^2(1)
+\sfrac{1}{3!} Y_{2x} \bar Q^2(x)\bigg\}\bar Q(x)
\nonumber\\
+Y_r \bigg\{ \bar Q(x) \!\int_x^1\!\!\d x'\,\bar R(x')
+\!\int_0^x\!\!\d x'\,\bar Q(x')\bar R(x')\bigg\} .
\ear
Differentiation with respect to $x$ results in
\be
\partial_x\bar Q(x)=\bar Y(x) \,\partial_x\bar Q(x)
\ee
with the same $\bar Y(x)$ as above. This allows a non trivial solution for 
$\bar Y(x)=1$. 

\subsection{Results in next to leading order} \label{O2}

In particular for $x=1$ and in second order we get
\bel{sbro}
\bar Q(1)=q_{EA}=\frac{1-Y_0}{Y_1}-\half\,\frac{(Y_2+Y_{2x})(1-Y_0)^2}{Y_1^3} .
\ee
Lowering the temperature a non zero Edwards-Anderson order parameter 
$q_{EA}>0$ shows up first at a critical temperature where
\be
Y_0=I^2K\,V_2^2=K\,\tanh^2(\beta I)=1 .
\ee
This determines the critical temperature
\be
T_c=\frac{I}{\artanh\big(1/\sqrt{K}\big)}
\ee
and choosing
\be
I=\artanh\big(1/\sqrt{K}\big)
\ee
the critical temperature is $T_c=1$.

Expanding the vertices and coefficients \req{hesd} to \req{kytf} in powers of 
$\epsilon=1-T$
\bear
Y_0=1+2 I \sfrac{K-1}{\sqrt{K}} \epsilon+I (I K-3 I+2\sqrt{K}) \sfrac{K-1}{K}
\epsilon^2,
\nonumber\\
Y_1=-2I^2(K-1)-4 I^2 \big(I \sqrt{K}+1\big)\big(K-1\big)\epsilon,
\nonumber\\
Y_2= 10 I^4 (K^2-1), \qquad
Y_{2x}=4 I^4 (K^2-1), \qquad
Y_r=2 I \sqrt{K}
\ear
and \req{sbro} becomes
\bel{jtge}
q_{EA}=\frac{1}{I\sqrt{K}}\,\epsilon
+\bigg[2 \frac{K+1}{K}-\frac{1}{I\sqrt{K}}\bigg] \,\epsilon^2 .
\ee
Differentiating \req{hqfc} with respect to $x$ gives
\bel{ksfe}
\bar R(x)=\frac{Y_{2x}}{Y_r}\,\bar Q(x)\,\partial_x\bar Q(x)
=2\,I^3\frac{K^2-1}{\sqrt{K}} \,\bar Q(x)\,\partial_x\bar Q(x) .
\ee
This verifies that $\bar R(x)\sim \epsilon^2$. 

In equilibrium a fluctuation-dissipation theorem (FDT) holds, i.e. $r(t)=-\beta \partial_t q(t)$. As pointed out earlier this is also expected to hold in the glassy state for $t<t^*$.
On the long time scale, however, the FDT is violated. We may introduce a measure $X$ for the violation of the FDT
\bel{lkte}
\bar r(\tau)=-\beta \,X(\tau)\,\partial_\tau \bar q(\tau), \qquad
\bar R(x)=\beta  \bar X\big(\bar Q(x)\big)\,\partial_x\bar Q(x)
\ee
with
\be
\bar X(q)=\frac{Y_{2x}}{Y_r}\,q=2\,I^3\frac{K^2-1}{\sqrt{K}} \,q \qquad
\mbox{for}\quad 0<q<q_{EA} .
\ee
in lowest order.

The fact that $\bar X$ depends on $\bar Q$ only holds in higher orders as well. 
For $t<t^*$ the correlation function obeys $q(t)>q_{EA}$ and we can set 
$\bar X(q)=1$ for $t<t^*$. This quantity is related to Parisi's overlap distribution function $P(q)=\partial_q \bar X(q)$. 

Eq.\req{ksfe} does not fix the actual form of $\bar Q(x)$ or $\bar R(x)$. Using the so called Parisi gauge \cite{DGO81}, $\beta X\big(\bar Q(x)\big)=x$, 
\be
\bar Q(x)=\frac{Y_r}{Y_{2x}} x \qquad \mbox{for} \quad x<x^*
\ee
with
\be
x^*=\frac{Y_{2x}}{Y_r}\, q_{EA}
\ee
and $t^*=t_w^{x^*}$.

For $K\to\infty$ the known results for the SK-model are recovered.

\section{Stability analysis}
\subsection{Solutions for general distribution of couplings}
For general distributions $P(J)$ the difference 
$\bar Y'(x)-\bar Y(x)$, given in \req{gilr} is non zero. In order $\epsilon$ the last term $\sim 1-I^2 K \overx{V_2^2}$ does not contribute. The remaining term contains
\be
\overx{V_2^2}\,\overx{V_2^3V_4}-\overx{V_2^4}\,\overx{V_2V_4}
\ge 0,
\ee
which is positive according to \req{ntas} unless $J=\pm I$. Accordingly  
$\bar Y'(x)>\bar Y(x)$ for general distributions $P(J)$.

This still allows for solution (a) of section \ref{FNU} with $\bar R(x)=0$. 
The last term in \req{hqfc} vanishes and $\bar Q(x)\!=\!q_{EA}$, given in \req{jtge}. This solution does not show any time dependence on long time scales. This property is shared by correlation functions of higher order, $C_{n,0}(\tau_1\cdots \tau_n)=C_{n,0}$. Correlation-response-functions $C_{n,m}(\cdots)$ with $m>0$ vanish. Reconstructing the full distribution 
${\cal Q}(\{\hat\kappa,\kappa\})$ from \req{ehmp} one finds the replica symmetric solution \cite{MP01}
\be
{\cal Q}(\{\hat\kappa,\kappa\})=\intx\d\bar\kappa\,P(\bar\kappa)\,
\e^{\int\!\d\sigma \hat\kappa(\sigma)[\kappa(\sigma)-\bar\kappa]}
\ee
with
\be
P(\kappa)=\prod_{i=1}^K \intx\d\kappa_i\,P(\kappa_i)\,
\overline{\delta\big(h+\sum_i u(J_i,\kappa_i)-\kappa)}^{J} .
\ee
In next to leading order
\bel{dvfe}
\bar Y(x)=Y_0+Y_1 \,q_{EA}+\big[\half Y_2+\sfrac{1}{3!}Y_{2x}\big]\,q_{EA}^2
\ee
which yields
\be
q_{EA}=\frac{1}{I\sqrt{K}}\,\epsilon
-\bigg[\frac{13}{3}\,\frac{K+1}{K}-\frac{1}{I \sqrt{K}}\bigg] \,\epsilon^2 .
\ee
As shown below this solution is unstable.

Solution (b) of section \ref{FNU} requires $\bar Q(x)=0$ and with \req{hqfc}
\be
\partial_x \bar Y(x)=Y_0+Y_r\!\int_x^1\!\!\d x'\,\bar R(x')=1 .
\ee
This has, however, no solution with $\bar R(x)\ne 0$.

\subsection{Stability analysis}
Various criteria can be used to test the stability of a given solution. One may for instance ask whether the correlation-response-functions on the short time scale approach the values determined by the dynamics on the long time scale. For example the pair correlation function $q(t)=\av{\sigma(t)\sigma(0)}$ should approach $q_{EA}$ for $t\approx t^*$. This decay is expected to be algebraic, but a corresponding analysis is outside the scope of 
this paper.

The general procedure used in statics as well as in the present formulation of dynamics consists in iteratively connecting subtrees to a new tree and expressing the properties at the new vertex by properties at the base vertices of the subtrees searching for fixed points of this mapping. A necessary condition for stability is the decay of small perturbations under this mapping. 
Eqs.\req{mfqa} and \req{rkda} are actually such mappings.

For $T>T_c$ the trivial solution $\bar Q(x)=0$ and $\bar R(x)=0$ is expected to be valid. A small perturbation $\delta \bar Q_i(x)$ on site $i$ is mapped onto
\be
\delta \bar Q_o(x)= I^2 K V_2^2 \delta \bar Q_i(x)
\ee
at site $o$, and a corresponding mapping for $\bar R_0(x)$. Since 
$I^2 K V_2^2<1$ for $T>T_c$ the trivial solution fulfills this stability criterion. By the same token this solution is unstable for $T<T_c$.

The stability criterion for solution (a) reads with \req{dvfe} and $\bar Y(x)=1$
\be
\delta \bar Q_o(x)= \big[1+\sfrac13 Y_{2x} \, q_{EA}^2 \big]\,
\delta \bar Q_i(x)>\delta \bar Q_i(x)
\ee
and the same for $\delta \bar R(x)$. This solution is therefore unstable for 
$T<T_c$ as well. 

Finally solution (c) for $J=\pm I$ is investigated. We may test the stability with respect to a perturbation $\delta \bar Q(x)$ and 
$\delta \bar R(x)=X\big(\bar Q(x)\big)\,\partial_x \delta\bar Q(x)$. This yields with $\bar Y(x)=1$, \ $\delta \bar Q_0(x)=\delta \bar Q_i(x)$ and 
$\delta \bar R_0(x)=\delta \bar R_i(x)$. This means that this solution is marginal with respect to perturbations of this kind. Investigating a perturbation $\delta \bar R(x)$ with $\delta \bar Q(x)=0$ one obtains
\be
\delta \bar R_0(x)=\delta \bar R_i(x)+Y_r \!\int_x^1\!\d x'\,\delta \bar R_i(x')
\,\bar R(x) .
\ee
Considering the second contribution a perturbation at some value $x'$ creates fluctuations at $x<x'$ only. This means that the perturbation at $x'$ is not enhanced due to this term and this solution is marginal with respect to a perturbation of this kind as well. Such a marginal stability criterion is actually expected because of the reparametrization invariance mentioned at the end of Section \ref{O2}.

\section{Discussion}

In this paper we have shown that the long time dynamics of an Ising spin glass with binary couplings $J=\pm I$ on a Bethe lattice is of the form known from the Sherrington-Kirkpatrick model. This is remarkable insofar as the interaction is restricted to nearest neighbors. In contrast to a lattice in finite dimensions, however, the typical size of closed loops scales with $\ln(N)$ for a system of $N$ sites.

The second main result of this paper, the failure of a corresponding solution for general distributions $P(J)$, is unexpected, and the nature of the low temperature phase in this case is not known. A similar breakdown of a replica or cavity method calculation for general $P(J)$ can not be excluded since the relevant equations have been evaluated for binary coupling only \cite{MP01}. It might be of interest to perform an expansion of the relevant equations around $T_c$, possibly within the extended replica scheme proposed by de Dominicis et.al \cite{DGO81}. In this scheme two order parameter functions corresponding to $\bar Q(x)$ and $\bar R(x)$ are used.\\
\ \\ 
This paper is dedicated to David Sherrington on the occasion of his 65th birthday.



\section*{References}
\begin{thebibliography}{99}

   

\bibitem{MPV87} 
	M. M\'ezard, G. Parisi and M.A. Virasoro, \textit{Spin Glass Theory and 		Beyond} (World Scientific, Singapore, 1987)
	
\bibitem{SCM04}  
	G. Semerjian,  L. F. Cugliandolo and A. Montanari
	Journal of Statistical Physics, \textbf{115}, 493, (2004)

\bibitem{SZ82}	
	H. Sompolinsky and A. Zippelius,Phys. Rev. B \textbf{25}, 6860 (1982)
	
\bibitem{Ho84}	
	H. Horner, Z. Phys. B \textbf{57}, 29, 39 (1984)

\bibitem{SK75}	
	D. Sherrington and S. Kirkpatrick, Phys. Rev. Lett. \textbf{35},
              1972 (1975)
	 	
\bibitem{KT87}	
	T. R. Kirkpatrick and D. Thirumalai, Phys. Rev. Lett. \textbf{58}, 2091 		(1987); 	Phys. Rev. B \textbf{36}, 5388 (1987) 

\bibitem{CHS93}	
	A. Crisanti, H. Horner and H.-J. Sommers, Z. Phys. B \textbf{92}, 257 (1993)
	
\bibitem{CK93}	
    L. F. Cugliandolo and J. Kurchan, Phys. Rev. Lett. \textbf{71}, 173 (1993)
	
\bibitem{FH90}	
	M. Freixa-Pascual and H. Horner, Z. Phys. B \textbf{80}, 95 (1990)
    
\bibitem{BBM96}  
A. Barrat, R. Burioni and M. Mezard, J. Phys. A \textbf{29}, L81 (1996)
	
\bibitem{Ho92}	
	H. Horner, Z. Phys. B \textbf{86}, 291 (1992)
	
\bibitem{Ho07}	
	H. Horner, arXiv:0707.2714, to be published in Eur. Phys. J. B (2008) 
	
\bibitem{MP01}	
	M. M\'ezard and G. Parisi, Eur. Phys. J. B \textbf{20}, 217, (2001)

\bibitem{Mo87}	
	P. Mottishaw, Europhys. Lett. \textbf{4}, 333 (1987)
	
\bibitem{DG89}	
C. De Dominicis, Y.Y. Goldschmidt, J. Phys. A \textbf{22}, L775 
(1989), Phys. Rev. B \textbf{41}, 2184 (1990)

\bibitem{ZKH00}	
W. Zippold, R. K\"uhn and H. Horner, Eur.Phys.J. B \textbf{13}, 531 (2000)

\bibitem{DGO81}  
C. De Dominicis,M. Gabay and H. Orland, J. Physique Lettres \textbf{42}, L523, (1981)  
	


\end{thebibliography}
\end{document}